\documentclass[manuscript]{emulateapj}
\shorttitle{PAPER 8-Station Results}
\shortauthors{Parsons, et al.}

\usepackage{amsmath}
\usepackage{graphicx}
\usepackage[figuresright]{rotating}
\usepackage{natbib}
\usepackage{pdflscape}

\begin{document}
\title{The Precision Array for Probing the Epoch of Reionization: 
8 Station Results}

\author{
Aaron R. Parsons\altaffilmark{1,2}, 
Donald C. Backer\altaffilmark{1}, 
Richard F. Bradley\altaffilmark{3,4,5},
James E. Aguirre\altaffilmark{6},
Erin E. Benoit\altaffilmark{5}, 
Chris L. Carilli\altaffilmark{7},
Griffin S. Foster\altaffilmark{1}, 
Nicole E. Gugliucci\altaffilmark{3}, 
David Herne\altaffilmark{8}, 
Daniel C. Jacobs\altaffilmark{6},
Mervyn J. Lynch\altaffilmark{8},
Jason R. Manley\altaffilmark{9,10}, 
Chaitali R. Parashare\altaffilmark{4}, 
Daniel J. Werthimer\altaffilmark{9,11},
Melvyn C.~H. Wright\altaffilmark{1}
}

\altaffiltext{1}{Astronomy Dept., U. California, Berkeley, CA}
\altaffiltext{2}{Pre-Doctoral Researcher, NAIC Arecibo Obs., Arecibo, PR}
\altaffiltext{3}{Astronomy Dept., U. Virginia, Charlottesville, VA}
\altaffiltext{4}{Dept. of Electrical and Computer Engineering, 
U. Virginia, Charlottesville, VA}
\altaffiltext{5}{National Radio Astronomy Obs., Charlottesville, VA}
\altaffiltext{6}{Dept. of Physics and Astronomy, U. 
Pennsylvania, Philadelphia, PA}
\altaffiltext{7}{National Radio Astronomy Obs., Socorro, NM}
\altaffiltext{8}{Dept. of Imaging and Applied Physics, 
Curtin U. Technology, Perth, Western Australia}
\altaffiltext{9}{Berkeley Wireless Research Cen., 
U. California, Berkeley, CA}
\altaffiltext{10}{Karoo Array Telescope, Cape Town, South Africa}
\altaffiltext{11}{Space Science Laboratory, U. California, 
Berkeley, CA}

\begin{abstract}
We are developing the Precision Array for Probing the Epoch of Reionization
(PAPER) to detect 21cm emission from the early Universe, when the first stars
and galaxies were forming.  We describe the overall experiment strategy and
architecture and summarize two PAPER deployments: a 4-antenna array in
the low-RFI environment of Western Australia and an 8-antenna array at our
prototyping site in Green Bank, WV.  From these activities we report on system
performance, including primary beam model verification, dependence of system
gain on ambient temperature, measurements of receiver and overall system
temperatures, and characterization of the RFI environment at each deployment
site.   

We present an all-sky map synthesized between 139 MHz and 174 MHz using data
from both arrays that reaches down to 80 mJy (4.9 K, for a beam size of 
2.15e-5 steradians at 154 MHz), with a 10 mJy (620 mK) thermal noise level that
indicates what would be achievable with better foreground subtraction.
We
calculate angular power spectra ($C_\ell$) in a cold patch and determine them
to be dominated by point sources, but with contributions from galactic 
synchrotron
emission at lower radio frequencies and angular wavemodes.  
Although
the cosmic variance of foregrounds dominates errors in these power spectra, we
measure a thermal noise level of 310 mK at $\ell=100$ for a 1.46-MHz band
centered at 164.5 MHz.  This sensitivity level is approximately three orders of
magnitude in temperature above the level of the fluctuations
in 21cm emission associated with reionization.
\end{abstract}


\section{Introduction}
\label{sec:intro}

The Epoch of Reionization (EoR) marks the transition of the primordial
intergalactic medium (IGM) from a neutral to a highly ionized state as a result
of radiation from the first stars and massive black holes \citep{loeb_barkana2001}.  This
phase transition represents a key benchmark in the history of cosmic structure
formation and a major frontier of cosmic evolution yet to be explored.
Observations of Gunn-Peterson (GP) absorption by the IGM towards
distant quasars \citep{becker_et_al2001,fan_et_al2006} and large-scale Cosmic Microwave
Background (CMB) polarization from Thompson scattering \citep{page_et_al2007}
have constrained cosmic reionization between redshifts $6<z<14$.  However,
deeper exploration of reionization via these probes faces fundamental 
limitations:
the GP-effect saturates at low neutral fractions and CMB polarization is an
integral measure of the Thompson optical depth to recombination. 

The most incisive probe of EoR is direct observation of the neutral IGM using
the hydrogen 21cm line \citep{furlanetto_et_al2006,barkana_loeb2005a}.  The rich
astrophysics traced by HI 
and the intrinsic three-dimensionality of the signal make this approach
especially appealing \citep{barkana_loeb2005b,loeb_zaldarriaga2004}.  However, 
the challenges of exploring reionization with red-shifted 21cm emission
in an observing band below 200 MHz are daunting.  EoR detection experiments require
unprecedented levels of instrumental calibration and foreground
characterization.  The brightness temperatures of polarized
galactic synchrotron emission, continuum point-sources, and
galactic/extra-galactic free-free emission can exceed the expected $\sim$10 mK 
fluctuations of the 21cm EoR signal by more than 5 orders of magnitude
\citep{zahn_et_al2007,santos_et_al2005}.  Wide fields-of-view (FoVs), large
fractional bandwidths, large numbers of antennas, significant RFI environments,
and ionospheric variation all present challenges for
next-generation low-frequency arrays.

The Precision Array to Probe the Epoch of Reionization (PAPER) is a
first-generation experiment focused on statistical EoR detection, but with
sufficient sky coverage and sensitivity to detect the very rare, largest-scale
structures formed at the end of reionization.  PAPER represents a
focused effort to overcome the substantial technical challenges posed by using
large, meter-wave interferometric arrays to detect the 21cm EoR signal.  

In \S\ref{sec:arch}, we present the principle deployments of the PAPER
instrument.  We then outline the architectures of the analog
(\S\ref{sec:analog}) and digital (\S\ref{sec:digital}) signal paths and
describe the calibration pipeline (\S\ref{sec:calibration}) that is applied to
the data.  Finally, we present observational results in \S\ref{sec:results}
that include an all-sky map and angular power spectra measured toward a colder
patch of the synchrotron sky.

\section{PAPER Deployments}
\label{sec:arch}

PAPER is being developed as a series of deployments of increasing scope to
address the instrumentation, calibration, and foreground characterization
challenges that must be surmounted in order to detect a 21cm signal from EoR.
By characterizing and optimizing each component in the array
with careful engineering, we hope to reduce the complexity of 
data calibration and analysis.  Our staged approach allows for
a systematic investigation of observational challenges, with a capacity for
adaptation as the characteristics of our instrument and of interfering 
foregrounds are better understood.

\begin{table}[htp]
\begin{center}
\caption{\bf PAPER Deployment Characteristics}
\label{tab:deploy}
\begin{tabular}{|ll|llll|}
\hline
& & PGB-4 & PWA-4 & PGB-8 & PWA-64 \\
\hline
\multicolumn{2}{|l|}{Deployment Date}
& 2004 & 2007 & 2006-8 & (2009)\\
$N_{\rm ant}$ & & 4 & 4 & 8 & 64 \\
$N_{\rm pol}$ & & 1 & 1 & 1 & 4 \\
$\Omega_{\rm B}$ &(str) & .96 & .96 & 0.43 & 0.43 \\
$\Delta\nu_{\rm corr}$ & (MHz)& 100 & 150 & 150 & 100 \\
$N_{\rm chan}$ & & 256 & 2048 & 2048 & $\ge2048$ \\
$\tau_{\rm int}$ & (s) & 32 & 7.16 & 14.32 & $\le7.16$ \\
$\tau_{\rm obs}$ & (days) & 1 & 3 & 14 & 90 \\
$d_{\rm max}$ & (m) & 100 & 150 & 300 & 600 \\
\hline
\end{tabular}
\end{center}
\end{table}

There have been three principal PAPER deployments, whose characteristics,
along with a planned 64 antenna deployment, are
outlined in Table \ref{tab:deploy}; $N_{\rm ant}$ is the number of antennas in
each deployment; $N_{\rm pol}$ is the number of polarization cross-multiples
computed in the correlator; $\Omega_{\rm B}$ denotes the solid angle of the
primary beam at 150 MHz; $\Delta\nu_{\rm corr}$ is the correlated bandwidth;
$N_{\rm chan}$ is the number of frequency channels computed over the correlated
bandwidth; $\tau_{\rm int}$ is the integration time per visibility; $\tau_{\rm
obs}$ is the longest continual operation; $d_{\rm max}$ is the approximate
maximum baseline length. In its initial
2004 deployment at the NRAO facilities in Green Bank\footnote{The National
Radio Astronomy Observatory (NRAO) is owned and operated by Associated
Universities, Inc. with funding from the National Science Foundation.} 
the PAPER instrument consisted of
four sleeved dipoles above planar ground-screens
arranged on an east-west line with integrated differential amplifiers.
This minimum-redundancy
array, dubbed PGB-4, had a longest baseline of 100 meters and provided initial
field experience and a means by which to evaluate our analog and digital
electronics.  PGB-4 established PAPER's basic architecture as that of a transit
array of zenith-pointing dipoles connected to a central correlator via 
fixed-length cables running above ground.

\begin{figure}\centering
    \includegraphics[width=.95\columnwidth]{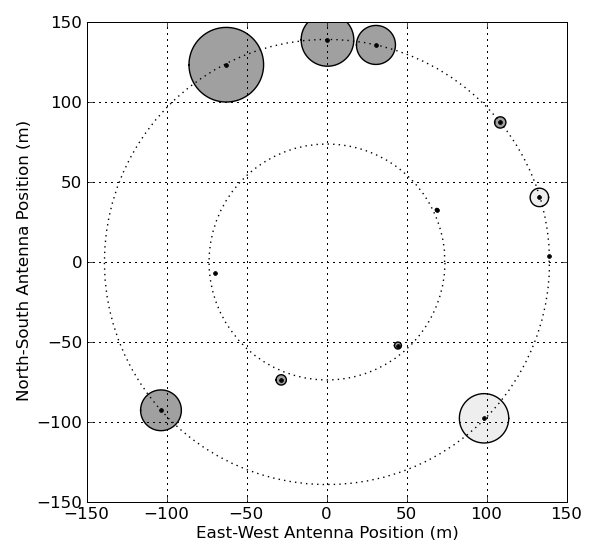}\caption{
Illustrated above are the antenna configurations, in topocentric coordinates,
for the PGB-8 deployment (outer, dotted) and the PWA-4 deployment (inner,
dotted).  Centered on each antenna are circles whose radii represent 
antenna elevations scaled by a factor of 10,
with light/dark gray indicating positive/negative elevation with respect to a
fiducial antenna.
    }\label{fig:antpos}
\end{figure}

The second PAPER deployment, PGB-8, began in 2006 and underwent continual
development until it was upgraded to PGB-16 in October 2008.
PGB-8 consisted of eight antennas deployed on a 300m
diameter circle (Fig. \ref{fig:antpos}) at the NRAO 
Green Bank Galford Meadow site, with a small hut
for rudimentary climate control built at the center of the array to house the
receivers and correlator.  As discussed in \S\ref{sec:analog}, 
antenna signals propagate to the hut over 75~ohm coaxial cable.
The correlator
itself went through several cycles of improvement paralleling the development
of the scalable correlator architecture described in \S\ref{sec:digital}.
 In its
final state, the PGB-8 correlator employed a packetized correlator (PaCo-8)
that processed 2048 spectral channels across 150 MHz of bandwidth, computing
all 4 Stokes parameters, although the
analog system only supported one polarization.  In early 2008, we improved
the forward gains of antennas by adding $45^\circ$ side reflectors to the
ground-screens.

Our third deployment, PWA-4, was near the Murchison Radio Observatory (MRO)
site\footnote{We acknowledge the Wajarri-Yamatji people of Australia as the
Native Title Claimants of the purposed MRO lands and we thank them for
allowing scientific activity on the site.} in Western Australia in July 2007.
This path-finding array gave us firsthand experience with the logistics of
deploying and operating an array at a very remote site.  The success
of this deployment gave us confidence 
in the low RFI levels and data quality that
can be expected from a much larger array at this site.  PWA-4
consisted of four antennas with
planar ground-screens (no side reflectors) arranged in a trapezoid pattern
with a maximum baseline of 150m.  The 4-input
``Pocket Correlator'' (PoCo) described in \S\ref{sec:digital} correlated
signals from these four elements.

A fourth deployment, PGB-16, is underway.  A total of sixteen sleeved dipole
antennas using ground-screens with side reflectors have been deployed at
Galford Meadow.  Initial operations are proceeding in single-polarization mode
using PaCo-8, but will evolve to dual-polarization with a PaCo-16 correlator.

\section{The Analog System}
\label{sec:analog}

\begin{figure*}\centering
    \includegraphics[width=6 in]{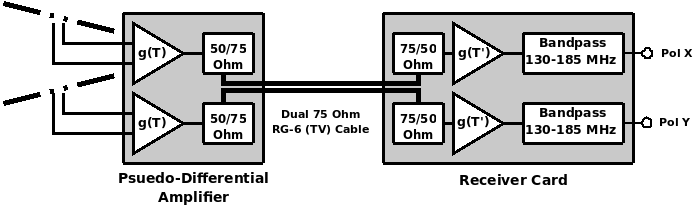}
    \caption{
The PAPER analog signal path flows from crossed dipole elements attached to a
Pseudo-Differential Amplifier (PDA), through RG-6 coaxial cable that runs over
the surface of the ground, to a Receiver Card (RC) that filters signals to a
130-MHz to 185-MHz band before transmitting them to the correlator.
Gain elements in both the PDA and RC are sensitive to ambient temperature.  An
enclosure is under development to maintain RCs at a constant 
temperature, so that the
only temperature-dependent gain that must be modeled is that of the PDA for
each antenna.
    }
    \label{fig:analog_arch}
\end{figure*}

The levels of instrumental calibration and foreground characterization that
will be required to model and remove signals interfering with an
EoR detection are unprecedented in
the 100-MHz to 200-MHz band expected to encompass reionization.  With this in
mind, we have taken care that each stage of our analog system (see Fig.
\ref{fig:analog_arch}) exhibits smooth responses as a function of
frequency and direction,
thereby minimizing the number of
parameters needed to describe these responses and limiting the magnitude 
of errors introduced by imperfect calibration.

\subsection{Antenna Design}
\label{sec:antenna}

\begin{figure}\centering
    \includegraphics[width=.95\columnwidth]{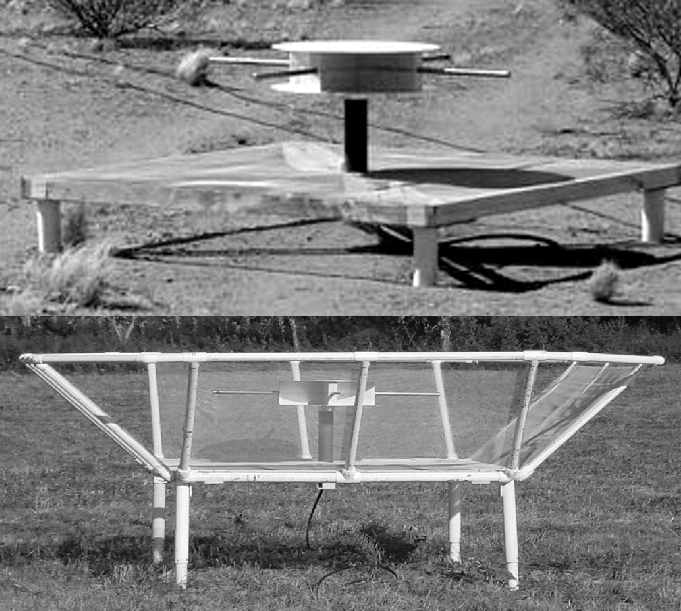}
    \caption{
PAPER antenna elements are dual-polarization sleeved dipoles mounted above
grounding structures.  These elements have been designed for smooth spatial and
spectral responses to facilitate calibration.  For the PWA-4 deployment
(top) this grounding structure was a flat, 2m$\times$2m  ground-screen mounted
on a wooden frame.  In the later PGB-8 deployment (bottom), this structure was
upgraded to include side reflectors that narrow the size of the primary beam to
more closely match the size of colder patches in the synchrotron sky.
    }
    \label{fig:ants}
\end{figure}

PAPER antennas are designed as rugged dual-polarization versions of the
sleeved dipole \citep{johnson1993}.  Crossed dipoles made from
copper tubing are encased between two thin aluminum disks, creating
a dual-resonance structure that broadens the antenna's frequency response.
The dimensions of the tubing and disks have been tuned for
efficient operation over a 120-MHz to 170-MHz band.
This design produces a spatially smooth primary beam pattern
that evolves slowly with frequency. 

The antenna design includes a grounding structure that alleviates the
gain variations that result from the effects of climatic conditions on
the dielectric properties of earth ground.
PGB-4 and PWA-4 deployments employed simple wire-mesh ground planes
supported by a wooden framework (Fig. \ref{fig:ants}, top).  This design was
improved to include planar wire-mesh reflectors that attach to the original
ground-screen (now supported by a steel framework) and rise outward
at a $45^\circ$ angle (Fig. \ref{fig:ants}, bottom), essentially becoming a
dual-polarization trough reflector \citep{arrl1988}.  This design 
produces a primary beam whose angular size
more closely matches the angular
size of colder patches in galactic synchrotron emission, effectively reducing
PAPER's sky-noise-dominated system temperature.  Reduction of horizon gain
also mitigates susceptibility to some RFI sources.  We have ensured that
these side-reflectors have not compromised the spatial
and spectral smoothness of the primary beam.

\begin{figure}
    \includegraphics[width=.95\columnwidth]{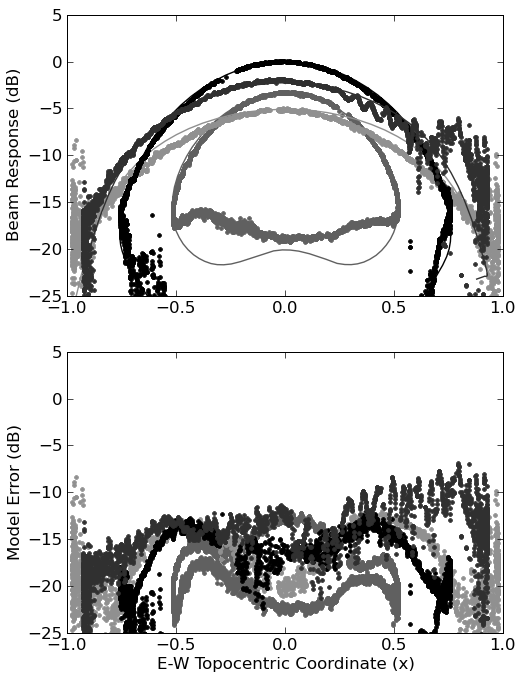}
    \caption{
Shown in the top plot are the predicted (lines) and measured (dots) responses 
of
the PGB-8 primary beam integrated between 138 MHz and 174 MHz towards 
(top to bottom at center of plot)
Cygnus A, Taurus A, Cassiopeia A, and Virgo A.  Responses are relative
to the primary beam's zenith response.  The lower plot shows the residual
response once a model of the primary beam has been subtracted.  Note that
the measured
beam response of Taurus A is
complicated by sidelobes of the Sun at $x>0.2$.
    }\label{fig:src_tracks}
\end{figure}

The smoothness of the primary beam as a function of frequency and direction
enable it to be effectively
parametrized using low-order spherical harmonics and frequency polynomials
according to the equation:
\begin{equation}
    a_\nu(\hat s) = \sum_{k=0}^7{\nu^k\left[\sum_{\ell=0}^8{
    \sum_{m=0}^\ell{a_{\ell m}(k)Y_{\ell m}(\hat s)}}\right]},
    \label{eq:ant_beam_pattern}
\end{equation}
where the frequency-dependence of the antenna response is modeled by a
seventh-order polynomial and the spatial variation of each polynomial
coefficient is described by $a_{\ell m}$ coefficients of low-order spherical
harmonic functions $Y_{\ell m}(\hat s)$.  This first-order beam model smoothly
interpolates to any chosen pointing and frequency, and has been assumed for all
of the analysis presented in this paper.  As shown in Figure
\ref{fig:src_tracks}, the model predicts the perceived flux densities of
sources with approximately 93\% accuracy for zenith-angles less
than $45^\circ$.  Errors in the beam model are primarily the result of
the spectral response of the primary beam being steeper as a function of
zenith angle than what is predicted by the model.
This
systematic effect is currently impeding the accurate measurement of spectral
indices of calibrator sources.

\subsection{Pseudo-Differential Amplifier Design}
\label{sec:balun_amp}

\begin{figure}\centering
    \includegraphics[width=.6\columnwidth]{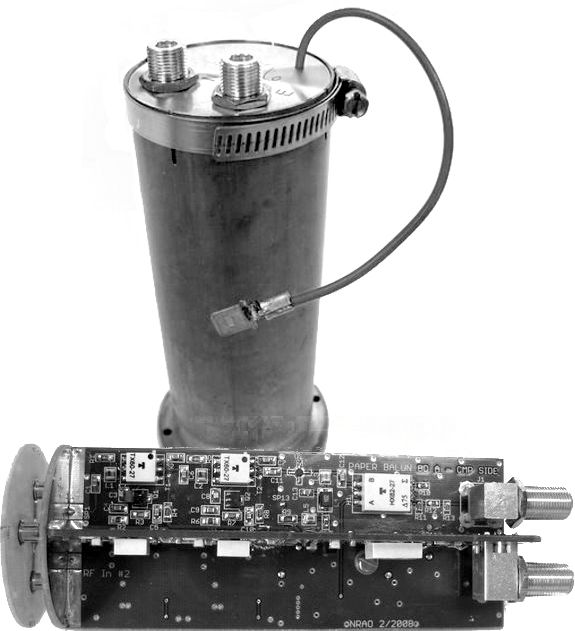}\caption{
The dual-polarization pseudo-differential amplifier, pictured in its casing
(upright) and with the casing removed (bottom), is housed in the riser
suspending the dipole elements above the ground screen.  The gain of this
amplifier is sensitive to temperature, with a coefficient of 
$-0.02 {\rm~dB}/{\rm K}$.
    } \label{fig:balun}
\end{figure}

A pseudo-differential amplifier (PDA, Fig. \ref{fig:balun}) is housed within
the riser suspending each dipole element above its ground-screen.  
Amplifiers employ three cascaded gain stages where the initial stage
consists of a differential amplifier based on NEC NE46100 bipolar junction
transistors.  This circuit features an inherent direct-current
pathway from input to ground, making the amplifier tolerant of
electrostatic charge build-up that can pose a serious problem in dry climates
\citep{jackson_farrell2006}.  The total $\sim$32 dB gain through the PDA unit
is sensitive to ambient temperature with a coefficient $H=-0.02 {\rm~dB}/{\rm
K}$ over the range 5--50$^\circ$C.  This coefficient enters into the voltage
gain $g_\nu$ of an antenna as
\begin{equation}
    g_\nu(t)=g_\nu(t_0)H\left(T(t)-T(t_0)\right),
    \label{eq:gain_vs_temp}
\end{equation}
where $T(t)$ is the ambient temperature as a function of time $t$ and $t_0$
represents a fixed time at which static gain calibration is performed.
Although not corrected for in the PWA-4 and PGB-8 systems, gain variation with
temperature can be partially offset by recording the average ambient
temperature of the array and applying a correction to all recorded
visibilities.  We are currently developing a system for using one antenna 
input as a
``gain-o-meter'' by swapping dipoles for a load attached at the front of
the PDA unit and using the gain variation of this input to normalize the gains
of the other antenna inputs.  A more advanced system could also record the
ambient temperature at each PDA to allow for per-antenna gain correction.

We have estimated the receiver temperature of the antenna/PDA system 
by modeling the frequency-dependent auto-correlation power $P_\nu$ as a
function of time $t$:
\begin{equation}
    P_\nu(t)=\left|g_\nu(t)\right|^2k_{\rm B}
        \left[T_{\rm sky}(t) + T_{\rm rx}\right]\sqrt{\Delta\nu}
\end{equation}
where $g_\nu(t)$ is defined in Equation \ref{eq:gain_vs_temp}; 
$T_{\rm sky}(t)$ is
the temperature of a model galactic synchrotron sky weighted by the primary
beam of an antenna; $T_{\rm rx}$ is the receiver temperature; $\Delta\nu$ is
the bandwidth of a channel.  $T_{\rm sky}$ was modeled by scaling a 408-MHz
sky map \citep{haslam_et_al1982} to 162 MHz using a spectral index of -2.52
\citep{rogers_bowman2008}.  As the galactic synchrotron sky rotates through
the primary beam, the perceived $T_{\rm sky}$ varies, providing a modulation
that enables the separation of sky and receiver temperatures. After 
applying a first-order correction for the temperature-dependence of amplifier 
gain using measured temperatures from a nearby weather station,
we fit an average
receiver temperature of 110 K for the PGB-8 system.

\subsection{Signal Transmission}

Coaxial cables transport antenna signals from each PDA unit to a central
processing location.  Because these
cables are not buried, they must be rugged to withstand harsh environmental
conditions and the occasional chew from local fauna.  We chose RG-6, 75~ohm 
cable
with a polyethylene jacket for its stable propagation characteristics as a
function of temperature and humidity and for its low cost.  Signal attenuation
over a 150m cable run is approximately 12 dB at 150 MHz, with a slope of
$+0.034 {\rm~dB}/{\rm MHz}$.  This cable also contains a wire suitable for
delivering DC power to PDA units.  

On the receiving end, dual-channel receiver boards consisting of amplifier
stages and band-limiting filters prepare signals for digitization in the
correlator.  The high gain of these receiver cards poses a significant
regenerative feedback concern not found in heterodyne systems
\citep{slurzberg_osterheld1961}.  This issue was mitigated by mounting the
receivers inside a special shielded enclosure \citep{bradley2006}.   The gain
of the amplifiers in the receiver cards is also sensitive to temperature, with
a coefficient of $-0.045 {\rm~dB}/{\rm K}$.  To avoid a second temperature
dependence in signal gain, a thermal enclosure with thermoelectric heat pumps
will be employed to stabilize the temperature of all receiver boards.  The
``gain-o-meter'' approach described in \S\ref{sec:balun_amp} 
will also provide a first-order correction.

\section{The Digital System}
\label{sec:digital}

A series of real-time digital FX correlators employing Field-Programmable Gate
Array (FPGA) processors addresses the growing digital signal processing 
(DSP) needs of visibility computation in PAPER deployments.
These correlators are based on the 
architecture described in \citet{parsons_et_al2008}, wherein DSP engines
transmit packetized data through 10-Gbit Ethernet (10-GbE) links to commercial
switches that are responsible for routing data between boards.  This
architecture, along with a set of analog-to-digital converters and modular
FPGA-based DSP hardware and a software environment for programming, debugging,
and running them, were developed in collaboration with the Center for
Astronomy Signal Processing and Electronics Research (CASPER, Parsons et al.
2006).
The flexibility
of this correlator design shortens development time, allowing a series of
correlators of increasing scale to be developed parallel to PAPER's incremental
build-out.

\subsection{CASPER Hardware and Gateware}

\begin{figure}\centering
    \includegraphics[width=.85\columnwidth]{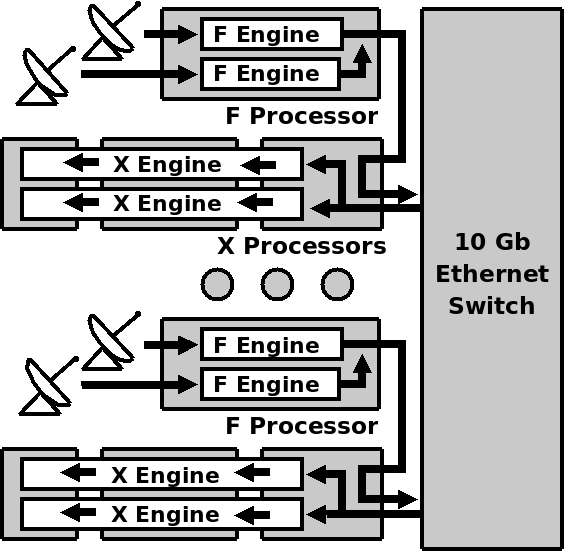}\caption{
PAPER's packetized correlator (PaCo) follows the frequency--cross-multiply (FX)
architecture developed by the Center for Astronomy
Signal Processing and Electronics Research (CASPER).  PaCo currently employs
IBOB boards for spectral processing (F engines, above) and
cross-multiplication in BEE2 boards (X engines, above); see
\citet{parsons_et_al2008} for details.  A 10-GbE switch is used to route data
between boards, so that data for a subset of channels from all antennas are
collected at each X engine, where all cross-multiples are computed.
    } \label{fig:paco_arch}
\end{figure}

The generic FX correlator architecture (Fig. \ref{fig:paco_arch})
on which PAPER correlators are based
consists of a set of modules responsible for digitizing,
down-converting, and channelizing antenna inputs (``F'' Engines),
followed by a set of signal processing modules that
cross-multiply all antenna and polarization samples for each frequency 
and accumulate the results (``X'' Engines).  
The problem of transmitting data from
every F engine to every X engine is solved by packetizing data according to the
10-GbE protocol and then using a commercial switch to sort data streams.
This approach, unique to the CASPER architecture for correlators of this size
and bandwidth, avoids custom backplanes and communication protocols that are 
tailored to
a single application, which are the traditional solution to the cross-connect 
problem in correlators.  

\begin{figure*}\centering
    \includegraphics[width=6 in]{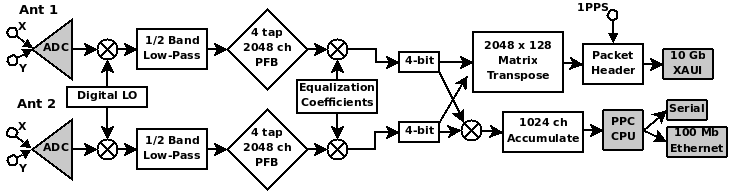}\caption{
The digital processing in an F engine of PAPER's packetized correlator (Fig.
\ref{fig:paco_arch}) also doubles as a stand-alone ``Pocket Correlator'' (PoCo)
for correlating four signal inputs.  In both designs, data is re-quantized to 4
bits after spectral decomposition.  To avoid signal distortion associated with
bit truncation, signals are equalized before truncation and the inverse
equalization is applied in post-processing.
    } \label{fig:poco_arch}
\end{figure*}

Signals from PAPER antennas are digitized by Atmel AT84AD001B
dual 8-bit ADC chips capable of digitizing two streams at 1 Gsample/sec.  While
overrated for this application, this chip has been used in many CASPER
applications and is extensively tested.  The wide bandwidth of this ADC
allows the entire PAPER bandwidth to be Nyquist-sampled as a single, real
voltage and then digitally mixed to baseband with perfect in-phase and
quadrature-phase components.  Two ADC boards connect to each of the CASPER IBOB
processing boards used as F engines in PAPER's correlator.  On the IBOB, one
FPGA is responsible for the signal flow outlined in Figure
\ref{fig:poco_arch}--namely, digitally down-converting four digitized signals,
spectrally decomposing the resulting baseband signals using a Polyphase
Filter Bank (PFB) \citep{crochiere_rabiner1983, vaidyanathan1990}, and
re-quantizing spectral data to 4 bits. 
PAPER correlators equalize antenna spectra before re-quantization
to ensure optimal linearity of the output power with respect
to input power,
as described in \S\ref{sec:gain_linearization}.  Output data
are passed briefly to the BEE2 board (described below) where they are formatted
into 10-GbE packets and transmitted to a commercial switch.  

A 12-port Fujitsu CX600 10-GbE switch is employed for routing packets
between F and X engines.  Each port on this switch carries data for
two antennas.  The performance of this switch was evaluated over a period of
16 hours, with pseudo-random data transmitted and received on each data link.
During this time, no errors were detected, placing a limit on the bit-error
rate of transmission at 2.2e-16 bits/s.  Furthermore, the asynchronous
correlator architecture employed by PAPER is tolerant of dropped packets, so
errors that do occur do not cause system failures.

We rely on CASPER's BEE2 boards for implementing X engine processing.  
Each of four 
FPGAs on a BEE2 operates independently to unscramble packets received 
from the 10-GbE switch.  Two X engine cores inside each BEE2 FPGA then
compute cross-multiplications for all $N$ antennas for 1/$N^{\rm th}$ of the
frequency channels and the results are accumulated for a selectable time
interval.  Output of accumulated data over a 100-Mbit Ethernet connection
currently imposes a 14-second minimum integration time.  We are currently
migrating this output
to 10-GbE, which will allow shorter integrations that more adequately
resolve ionospheric fluctuations.

\subsection{PoCo and PaCo-8}

\begin{figure}\centering
    \includegraphics[width=.85\columnwidth]{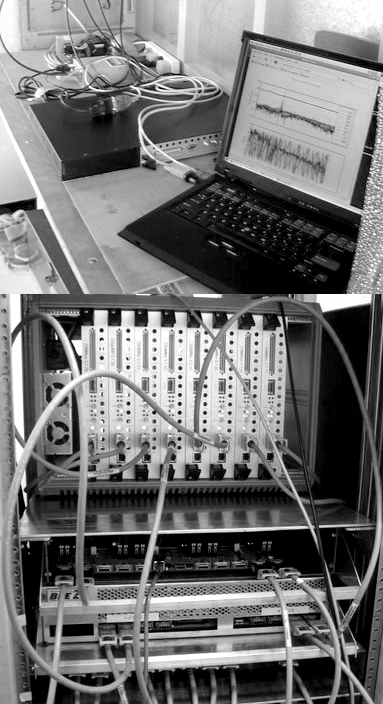}\caption{
For the PWA-4 deployment, a ``Pocket Correlator'' (top)
was employed for correlating four single-polarization
antenna inputs.  Data were recorded in MIRIAD format on a USB disk attached to
a laptop.  For PGB-8 and larger deployments, a packetized correlator (bottom)
connects multiple F engine processors through a
10-GbE switch to X engine processors and output data are collected by a server
for recording.
    } \label{fig:poco_paco_pics}
\end{figure}

The ``Pocket Correlator'' (PoCo, Fig. \ref{fig:poco_paco_pics}, top) used
with the PWA-4 deployment is a single IBOB system for correlating and
accumulating four input signals (Fig. \ref{fig:poco_arch}).  ADCs sample
each analog input at 600 MHz and a 75-MHz to 225-MHz
band is extracted digitally.  A 4-tap PFB decomposes this sub-band into 
2048 channels, which are then equalized to remove per-channel scaling 
differences before being
re-quantized to 4 bits.  Limited buffer space on the IBOB permits only 1024
channels (selectable from within the 2048) to be accumulated.  Accumulated
visibilities are output via serial connection to a host computer after being
integrated for 7.16 seconds.

The 8-antenna packetized correlator (PaCo-8, Fig. \ref{fig:poco_paco_pics},
bottom) used with the PGB-8 deployment employs four IBOBs and one BEE2 board
that communicate through a Fujitsu XG700 10-GbE switch, 
following the architecture outlined
in Figure \ref{fig:paco_arch}.  Each IBOB in this system operates identically
to PoCo, but branches data from the equalization module to a
matrix transposer for forming frequency-based packets.  Packet data for each
antenna are multiplexed through a point-to-point connection to a BEE2 FPGA and
then relayed in 10-GbE format to the switch.  A central CPU on each BEE2 board
collects all visibility data after it has been accumulated for 14.32 seconds
and transmits it over 100 Mb Ethernet to a server where it is written to disk
in MIRIAD format \citep{sault_et_al1995}.  In field deployments, correlators
are housed inside a special shielded enclosure \citep{bradley2006} to mitigate
self-interference.

\section{The Calibration Pipeline}
\label{sec:calibration}

The greatest challenge of using low-frequency interferometry to detect the
cosmic reionization lies in characterizing the celestial sky and
modeling the system response such that foregrounds to the EoR signal can be
effectively suppressed.  Various methods for estimating the effects of
instrumental calibration on a statistical EoR detection have been analyzed
\citep{morales2005,morales_et_al2006b,bowman_et_al2006}, but improving the
calibration and stability in early instruments remains an open problem
\citep{yatawatta_et_al2008,bowman_et_al2007a}.  Wide fields-of-view, large fractional
bandwidths, strong RFI environments, and ionospheric variation all complicate
the calibration of next-generation low-frequency arrays such as PAPER, the
Murchison Widefield Array\footnote{http://haystack.mit.edu/ast/arrays/mwa}
(MWA), the Long Wavelength Array\footnote{http://lwa.unm.edu} (LWA), and the
LOw Frequency ARray\footnote{http://www.lofar.org} (LOFAR).  Obtaining an
accurate sky model requires accurate models of primary beams, receiver
passbands, gain variation, and array geometry.  However, characterization of
these often depends on the availability of an accurate sky model.  The PAPER
approach to this problem has been to emphasize precision, with the goal of
characterizing system components to within 1\% in order to facilitate further
model refinement via sky modeling and self-calibration.

Although many standard tools exist for calibrating interferometric data,
we are developing an open-source software project
called Astronomical Interferometry in
PYthon\footnote{http://pypi.python.org/pypi/aipy} (AIPY) that modularizes
interferometric data reduction to facilitate the
exploration of new algorithms and calibration techniques.  The niche targeted
by this package are developing arrays whose evolving instrumental
characteristics require flexible, adaptable tools.  AIPY uses the dynamically
interpreted Python programming language to wrap together interfaces to 
packages such as MIRIAD \citep{sault_et_al1995} and HEALPix
\citep{gorski_et_al2005} using a common numerical array interface.  AIPY also
includes object-oriented implementations of a wide range of
algorithms useful for radio interferometry, including W-Projection
\citep{cornwell_et_al2003}, CLEAN \citep{hogbom1974}, Maximum-Entropy
deconvolution \citep{cornwell_evans1985,sault1990}, aperture synthesis imaging,
faceted map-making, visibility simulation, and coordinate
transformations.  AIPY relies heavily on 3rd party, public-domain
modules for 
scientific computing, allowing its development to focus on
interferometry-specific functionality.

AIPY analysis centers around parametrizing 
interferometric measurement equations for the instrument in question.  
However, before data from a PAPER correlator 
is ready to be compared
to the measurement equation shown in Equation \ref{eq:meas_eq},
it requires
gain linearization, RFI excision, and crosstalk removal.

\subsection{Gain Linearization}
\label{sec:gain_linearization}

Mitigating the effects of data quantization in the PAPER correlator requires
that power levels be carefully set for optimal SNR and
that output
data be converted to a linear power scale (see Ch. 8 of 
\citet{thompson_et_al2001}).  Correction factors for
gain non-linearities in digital systems employing 1-bit and 2-bit
quantization of voltage samples are well-known.  The non-linear power response
of the PAPER correlator, with its 4-bit correlation, is still significant. Our
approach to gain linearization is detailed in \citet{parsons_et_al2008}.
Inside the correlator, optimal SNR during re-quantization is ensured by applying
an equalization function that shifts data samples to the region of maximal
linearity in the 4-bit quantization response curve.  This equalization function
is updated hourly to adjust for changing mean power levels as the
galactic synchrotron sky traverses the dipole beam.  The first steps of PAPER
calibration are to deduce an average 4-bit value from the accumulated output of
the correlator, to invert the known quantization response for this value (Fig.
13 in \citet{parsons_et_al2008}), and then to remove the equalization function
used in the correlator.

\subsection{RFI Excision}

\begin{figure}\centering
    \includegraphics[width=.95\columnwidth]{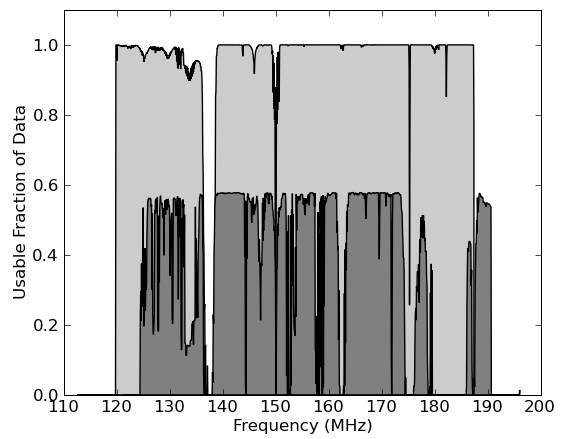}\caption{
Above are illustrated the fraction of usable integrations after flaggin 
RFI for each frequency channel over 3 days of observation with PWA-4
(light) and PGB-8 (dark).  Approximately 40\% of PGB-8 integrations are
unusable owing to saturation of ADC inputs by strong RFI.  The RFI environment
at PWA is exquisite, with fewer channels rendered unusable by continuous
narrow-band transmissions and almost no saturation events.
    } \label{fig:rfi_usable}
\end{figure}

As a general solution to identifying data points that depart from an
otherwise smooth function, we employ an algorithm that iteratively fits a
smooth polynomial to visibility amplitudes, subtracts that polynomial to
identify and exclude outlying points, and repeats.  This algorithm identifies
outlying points
by finding the mean and standard deviation of the logarithm of all
data points and defining a threshold in units of the standard deviation.  The
effectiveness of this technique is improved by first subtracting simulated
visibilities using accurate models of the array response and the celestial sky. 

This RFI excision algorithm is applied to spectra of cross-correlation data to
identify narrow-band interference and to time-series in auto-correlation data
to identify integrations where strong RFI such as aircraft communication
saturates the ADC input.  Such saturation events result in wide-band
``drop-outs'' where signal is lost across all frequencies.  In both cases,
detection of an interference event in a given channel/integration on any
baseline results in that data being flagged for all baselines.  Figure
\ref{fig:rfi_usable} illustrates the usable fraction of data averaged over a
full day for the PGB-8 and PWA-4 deployments.  The RFI environment in Western
Australia represents a marked improvement over that of Green Bank, with
fewer channels occupied by continuous RFI and almost no ADC saturation events.
Such events cause
approximately 40\% of PGB-8 integrations to be unusable.  

\subsection{Crosstalk Removal}

The inadvertent introduction of a correlation between two independent inputs
via a cross-coupling pathway is unavoidable in radio interferometers.  Many
systems use phase switching to help remove crosstalk introduced along the
signal path from an antenna to the ADC.  PAPER does not currently implement
phase switching, although we are exploring options for doing so.  However, in
the absence of (or in addition to) phase switching, there are other techniques
for removing crosstalk from visibility data, provided that such crosstalk
is adequately
stable with time.  PAPER's crosstalk is modeled as an additive component of 
measured visibilities that varies slowly on the timescale of several hours.
Because this crosstalk is stable on a timescale much longer than the inverse of
the typical fringe-rate of a source over the shortest baseline and because its
power is concentrated at low delays, PAPER crosstalk is easily suppressed using
frequency-based time-averages or delay/delay-rate filters
\citep{parsons_backer2009}.  As with RFI, crosstalk removal is
facilitated by the subtraction of simulated visibilities.

\subsection{PAPER's Measurement Equation}

At the heart of the PAPER analysis model is a measurement equation for the
single polarization visibility response ($V_{\nu}(t)$) of a baseline
at channel center frequency $\nu$ and time $t$:
\begin{equation}
    V_{\nu}(t)=G_{\nu}(t)\sum_n{A_\nu(\hat s_n)
    S_\nu(\hat s_n)e^{2\pi i\nu(\vec b\cdot\hat s_n + \tau) + \phi}}
\label{eq:meas_eq}
\end{equation}
where $\hat s_n$ is a  time-dependent unit vector pointing in the direction of
a source $n$ whose flux density is $S$; $G_\nu(t)\equiv
g_{i\nu}(t)g_{j\nu}^*(t)$ is the frequency-dependent baseline response, which
is expressed in terms of the voltage gain of each antenna ($g_{ij,\nu}$; Eq.
\ref{eq:gain_vs_temp}) for constituent antennas indexed by $i,j$;
$A_\nu(\hat s_n)\equiv a_{i\nu}(\hat s_n)a_{j\nu}^*(\hat s_n)$ is the
direction-dependent response of a baseline's primary beam, expressed in terms
of the electric field response of each antenna ($a_{ij,\nu}$; Eq.
\ref{eq:ant_beam_pattern}); $\vec b$ represents the baseline vector separating
antenna positions in units of light propagation time; $\tau$ is the relative
electronic delay of antenna signals into the correlator; and $\phi$ is a 
relative instrumental phase between antenna signals.

The design of our antenna and analog electronics leads to antenna-based gains
($a_{ij},g_{ij}$) that can be modeled by smooth functions in time 
and frequency with a
modest number of parameters.  The degrees of freedom in these parameters are
over-constrained by wide-band visibility data, even when using a sky model that
includes only a small number of point sources.  However, the larger FoV of
PAPER elements has complicated early calibration by decreasing the extent to
which a single source dominates the correlated signal between antennas.
Without isolation of sources, self-calibration cannot proceed as a direct
computation using raw data, but rather must proceed by fitting 
models of the array and the observed sky to remove baseline-dependent
interference patterns \citep{cornwell_fomalont1989}.  To address this problem,
we employ a two-tiered calibration process wherein initial coarse calibration
that employs source-isolation techniques is followed by multi-source,
least-squares fitting to visibilities for the parametrization described in
(\ref{eq:meas_eq}).

\subsection{First-Order, Time-Independent Self-Calibration}
\label{sec:self_calibration}

Phase calibration involves solving for the position and electronic signal delay
associated with each antenna.  The parameter space associated with phase
calibration contains many local minima separated by wavelength increments
projected toward calibrator sources.  To speed the phase calibration process,
we rely on theodolite-surveyed antenna positions.  When initial antenna
positions include errors greater than a wavelength, we have developed a
procedure for fitting positions that is reasonably robust against converging to
non-global minima.  First, a single source is isolated in data either by
selecting a time when one source is dominant, or by using coarse
delay/delay-rate filtering \citep{parsons_backer2009}.  Baseline components
$b_x,b_y$ and a throw-away phase term are then fit for a pair of nearby
antennas using a narrow set of frequency channels and a range of time over
which fringes wrap several times.  Next, data including a few other sources are
used to separate $b_z$ out of the remaining phase term.  As a final step in
single-baseline phase calibration, wide-bandwidth data are used to account for
the final phase as a combination of electronic signal delay ($\tau$) and a
small, frequency-independent phase term ($\phi$).

Following accurate phase calibration for a subset of antennas, another antenna
is added and the procedure outlined above is repeated using data for all
baselines that connect this antenna to the calibrated subset.  With phase
calibration for four antennas, preliminary antenna-based gain calibration is
possible.  PAPER visibilities have been flux-calibrated using Cygnus A with a
value of 1.09e4 Jy at 150 MHz and a spectral index of $-$0.69, which are
derived from the data in
\citet{baars_et_al1977}.  The beam model
described in \S\ref{sec:antenna} provided a first-order correction for 
the effects of the primary
beam on the Cygnus A spectrum.  In addition to an initial gain model that
incorporated the measured responses of the analog and digital filters in the
PAPER pipeline, fourth-order polynomials in frequency were fit for the bandpass
functions ($g_{i\nu}$) of antennas in the array.  

At each stage, the residuals remaining after flux-calibration can help to
highlight baselines with poorer fit scores--a hallmark of convergence to a
non-global minimum.  Once all problematic baselines have been eliminated,
simultaneous fitting of phase and gain parameters can proceed until
time-variable effects such as ionospheric distortion and changing primary beam
response towards strong sources dominate residual visibilities.

\subsection{Excision of Strong Point-Sources}
\label{sec:rm_pnt_srcs}

Strong point-sources, while useful for initial calibration, complicate further
calibration by degrading sensitivity toward other positions on the sky.  This
is especially true for small arrays, since the sidelobe response toward a
source scales approximately as $\sqrt{N_{vis}}$, where $N_{vis}$ is the number
of independent locations in the $uv$-plane
that are being phased and summed.  For PWA-4 and PGB-8, sources stronger than a
few hundred Jy can dominate the response of a synthesized beam, even when
using earth rotation synthesis.

\begin{figure*}\centering
    \includegraphics[width=7in]{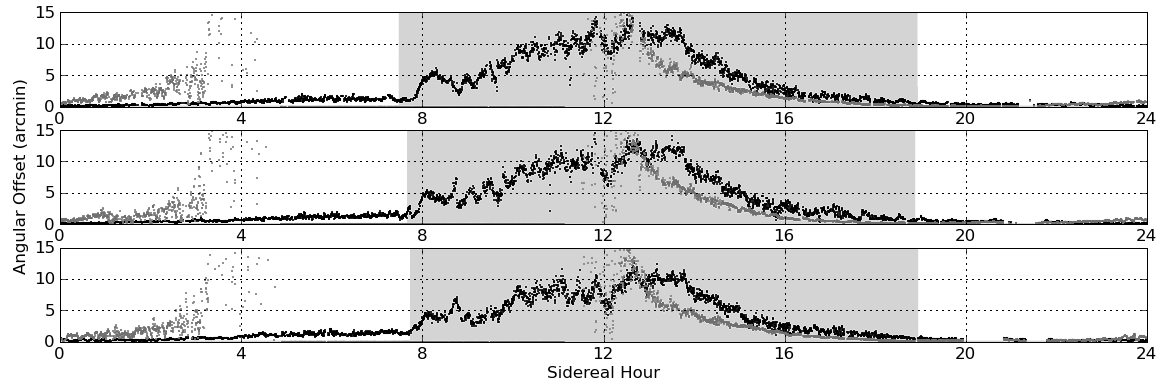}\caption{
Illustrated above are the magnitudes of the angular offsets, primarily in
the zenith direction, of Cygnus A (gray
dots) and Cassiopeia A (black dots) fit as a function of time
relative to their cataloged positions
for 3 days of observations (top to bottom) with PGB-8.
Both sources exhibit greater variability in position when the Sun was below the
horizon (gray shaded area), presumably as a result of greater ionospheric
variability at those times.  The offsets of both sources also exhibit a
smoothly varying component that repeats day-to-day, which may be attributed to
refraction from spherical and wedge components of the ionosphere 
\citep{komesaroff1960}.
    } \label{fig:ionosphere}
\end{figure*}

Deep removal of the strongest sources is essential for improving primary
beam models using sources at a variety of declinations. A time-invariant model
of source position, flux density, and spectral index is moderately effective at
removing these sources, depending on the accuracy of current beam models and 
the
activity of the ionosphere.  For PGB-8, we were able to achieve $\sim$95\%
suppression of sources using static modeling.  However, the residual flux
densities of the strongest five sources (Sun, Cygnus A, Cassiopeia A, Virgo A,
Taurus A) were still strong enough to introduce noticeable imaging sidelobes.
Improving source removal required fitting for offsets in source position
and flux spectrum as a function of time.  We attribute these variations to
ionospheric refraction (Fig. \ref{fig:ionosphere}) and defects in the primary
beam model, respectively.  All of these components were necessary to achieve
source suppression at a level of 1e-3 to 2e-4, depending on a source's position
within the primary beam.

\section{Observational Results}
\label{sec:results}

Results were obtained using 3-day observations with a single east-west linear
polarization for both PWA-4 and PGB-8 deployments.  For both sets
of data, the five strongest sources visible from the northern hemisphere (Sun,
Cygnus A, Cassiopeia A, Virgo A, and Taurus A) were removed following the
procedure described in \S\ref{sec:rm_pnt_srcs}.  Residuals associated with
these sources reflect deviations of their spectra (as viewed through an
imperfectly calibrated primary beam) from a strict power-law.  The Sun was
further suppressed via the application of a delay/fringe-rate filter
(\S\ref{sec:self_calibration}).  Although these filters potentially corrupt
wider areas of the map, the strength and variability of the Sun is such that
their application greatly improves imaging of the daytime sky.  

\subsection{PWA-4/PGB-8 All-Sky Map}
\label{sec:sky_map}

The all-sky map in Figure \ref{fig:allsky_map} illustrates average flux density
per beam area over a band between 138.8 MHz and 174.0 MHz. The northern
hemisphere was imaged using PGB-8 data with peak beam response at declination
$+38.5^\circ$.  The southern hemisphere was imaged with data from PWA-4, which
employed both a more compact array configuration and a broader primary beam.
Peak beam response in the southern hemisphere lies at declination
$-26.7^\circ$. Imaging over the full 35.2-MHz bandwidth was performed by
generating full maps in 1.46-MHz intervals and then summing maps.  Each map
was generated using 200 facets, with phase centers equally spaced around the
sphere.  While each imaged facet employed W-projection
\citep{cornwell_et_al2003} to correct for the curvature of the sky, facet
imaging was nonetheless required to incorporate differences in data weighting
between phase centers.  This weighting optimizes SNR by accounting for the
array's direction-dependent gain toward that location as it drifts through the
primary beam.

Thermal noise and point-source sidelobes both contribute to the noise level
seen in this all-sky map.  As mentioned in \S\ref{sec:results}, only the
strongest five sources have been removed from this map prior to imaging.
Image-domain CLEAN deconvolution was used to achieve modest suppression of
sidelobes from other sources.  The complexities of wide-field imaging with
drift-scan data make further post-imaging deconvolution impractical, since
CLEAN components of gridded visibilities may not be directly applied to
ungridded visibility data.  Our plan is to use imaging as a pathway for
identifying sources that may then be fit and removed from raw visibilities.
With this in mind, we have devoted little additional effort toward
suppressing sidelobes of imaged sources.  From the map in Figure
\ref{fig:allsky_map}, we measure an RMS flux density of noise between point
sources of 80 mJy (4.9 K using a synthesized beam area of 2.15e-5 steradians at
156.4 MHz).  As we discuss in \S\ref{sec:px_noise} and in
\S\ref{sec:pspec}, this noise level exceeds thermal noise by a factor of eight,
and is dominated by point-source sidelobes, with only a small contribution from
galactic synchrotron emission at lower frequencies.  
 
The position-dependent response of primary beams in each array is evidenced by
changing noise levels in the map as a function of declination.  Noise levels in
the map also change with right-ascension because system temperature depends on
the average temperature of galactic synchrotron emission across the primary
beam.  To illustrate the levels of thermal noise in the all-sky map, we
repeated the map-making process detailed above, but with alternating
integrations added with opposite signs, so that celestial sources with slow
time variation were heavily suppressed relative to thermal noise.  The map
generated by this technique, illustrated in Fig. \ref{fig:nos_map}, shows that
for the PGB-8 portion of the map, noise levels range from 10 mJy to 50 mJy (620
mK to 3.1 K using a synthesized beam area of 2.15e-5 steradians at 156.4 MHz)
in the declination range of peak sensitivity of the array.  

%

\subsection{Thermal Noise Contributions to PGB-8 Map}
\label{sec:px_noise}

The map statistics in the direction of a colder patch of the galactic
synchrotron were evaluated using 1.46-MHz facet images with phase centers at
$(11^h20,30^\circ00)$, $(11^h20,40^\circ00)$,
$(12^h00,30^\circ00)$, and $(12^h00,40^\circ00)$.  Each frequency interval and
phase center was imaged twice with independent sets of alternating (odd and
even time index) 14-second integrations from the 3-day PGB-8 observation
described in \ref{sec:sky_map}.  For each facet, these two images were summed
and differenced to produce images of the celestial sky and of thermal noise,
respectively.  For the following analysis, image statistics were evaluated
independently over $10^\circ$ diameter circular patches around each phase
center and averaged over facets afterward.

\begin{figure}\centering
    \includegraphics[width=.95\columnwidth]{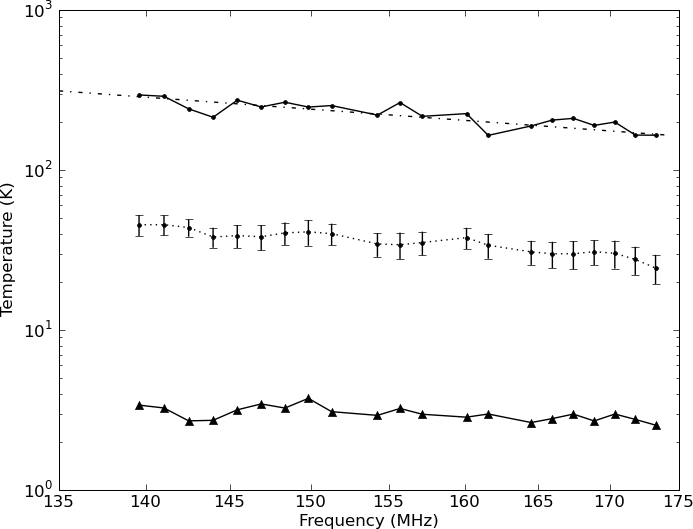}
    \caption{
We imaged a colder patch of synchrotron sky
using 3 days of PGB-8 data grouped into 1.46-MHz bands,
adding integrations with alternating signs so that
thermal noise could be estimated from RMS pixel values
(solid, triangles).  From these values, a system temperature
was inferred (solid, dots) using Equation \ref{eq:px_temp}.  Comparison with a
predicted synchrotron spectrum for this patch of sky (dot-dashed) indicates
that sky-noise dominates the system temperature of PGB-8.  The average sky
temperature (dotted, $2\sigma$ error bars) measured over these facets is
substantially lower than the predicted synchrotron spectrum because large-scale
structure is resolved out by the partially filled aperture of PGB-8.
    } \label{fig:tsys_vs_freq}
\end{figure}

Thermal noise levels were inferred from 
measured RMS pixel values according to the equation:
\begin{equation}
    \sigma_{\rm px} = \left<\frac{2k_BT_{\rm rms}\Omega_s}{\lambda^2}\right>
    =\frac{2k_BT_{\rm sys}\Omega_B}{\lambda^2
    \sqrt{\Delta\nu\tau N(N-1)}},
    \label{eq:px_temp}
\end{equation}
where $\Omega_s$ is the solid angle of a synthesized beam; $\Omega_B$ is the
solid angle of the primary beam; $\Delta\nu$ is the bandwidth of data used in
the image; $\tau$ is the integration time; $N$ is the number of antennas in
the array; and $<...>$ denotes the RMS value.  The thermal noise images 
produced by differencing integrations
yield the $T_{\rm rms}$ estimates plotted in Fig. \ref{fig:tsys_vs_freq} for
each 1.46-MHz band. 

From each measured $T_{\rm rms}$ we also inferred the system temperature
$T_{\rm sys}$ present before integration in the correlator.  A comparison of 
this
temperature with the $T_{\rm sky}$ predicted for a colder patch of the galactic
synchrotron (see \S\ref{sec:balun_amp}) shows that the PGB-8 system temperature
is dominated by sky-noise.  Finally, the standard deviation of pixel values in
the summed facet images exceeds thermal noise, implying
contributions from celestial point sources, their sidelobes, and galactic
synchrotron emission.  The temperature associated with this pixel distribution
reflects a sky temperature as perceived by the partially filled aperture of
PGB-8. This perceived temperature is approximately a factor of 6 lower than the
temperature of the galactic synchrotron because sparse samplings of short
$uv$-spacings cause much of the emission from this source to be resolved out.

\subsection{Angular Power Spectra}
\label{sec:pspec}

\begin{figure}\centering
    \includegraphics[width=.95\columnwidth]{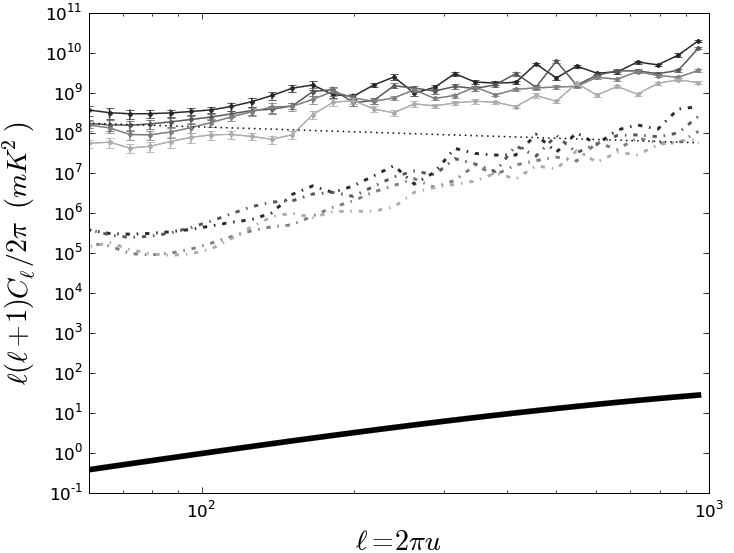}
    \caption{
Shown above are angular power spectra measured at 146.9, 155.7, 164.5, and
173.3 MHz (solid, top to bottom at $\ell=100$, respectively, with $2\sigma$
error bars), averaged over 1.46-MHz intervals for four $10^\circ$ facets near
$11^h40,35^\circ00$.  These spectra contain contributions from both point
sources and galactic synchrotron emission. Dominant errors are from cosmic
variance, which has been estimated from the standard deviation of measurements
at the same $\ell$ in the $uv$-plane.  Dot-dashed lines indicate angular power
spectra of ``noise images'', generated by adding integrations with alternating
signs, and are indicative of the thermal noise level in these measurements.
Also plotted: a fiducial reionization power spectrum at $z=9.2$ ($\nu=$140 MHz,
thick solid) and a predicted galactic synchrotron spectrum at 147 MHz (dotted)
\citep{tegmark_et_al2000,santos_et_al2005}.
    } \label{fig:Cl_vs_l}
\end{figure}

\begin{sidewaysfigure*}[p]\centering
    \includegraphics[width=9.5in]{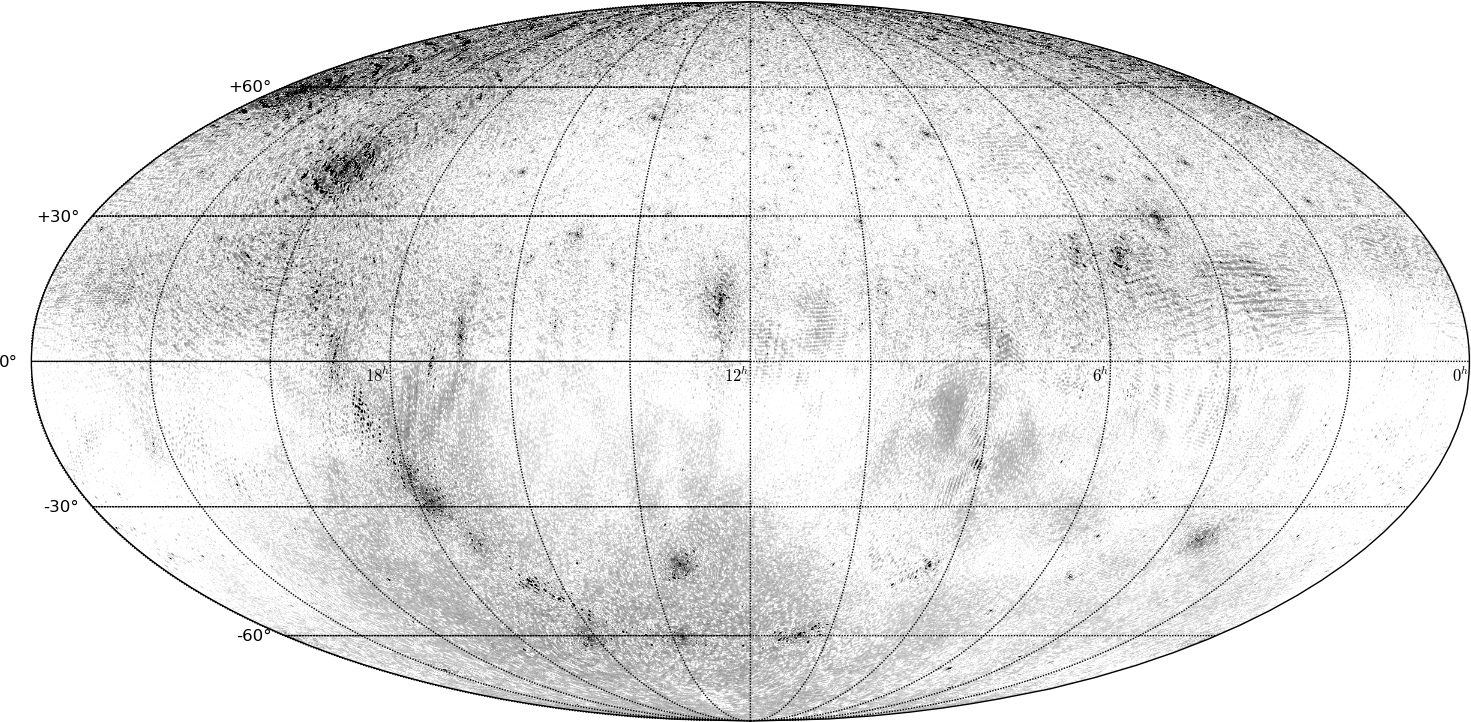}
    \caption{
Illustrated above is an all-sky equatorial (J2000) map (see
\S\ref{sec:sky_map}) averaged between 138.8 MHz and 174.0 MHz, in units of
${\rm log}_{10}(Jy/Beam)$ ranging from $-1$ (white) to $1$ (black).  Northern
hemisphere data were obtained with data from PGB-8 and southern
hemisphere with data from PWA-4, both with 3 days of observation
using a single east-west linear polarization.  Models of Cygnus A
$(19^h59,40^\circ44)$, Cassiopeia A $(23^h23,58^\circ49)$, Taurus A
$(05^h35,22^\circ01)$, Virgo A $(12^h31,12^\circ23)$, and the Sun
$(1^h10,7^\circ27)$ have been subtracted/filtered from visibilities prior to
imaging.  In the northern hemisphere, SNR peaks near $(12^h00,40^\circ00)$;
sensitivity degrades with declination because of declining primary beam
response and with right ascension because of increasing levels of sky-noise
from galactic synchrotron emission.  In the region of peak SNR, measured
temperatures reach a minimum of 4.9 K, as measured by the incompletely sampled
aperture of PGB-8.  Thermal noise in this same area is measured to be 620 mK
(see Fig. \ref{fig:nos_map}), indicating that point-source sidelobes and
diffuse galactic synchrotron emission dominate the noise-floor in this map
}
    \label{fig:allsky_map}
\end{sidewaysfigure*}

\begin{sidewaysfigure*}[p]\centering
    \includegraphics[width=9.5in]{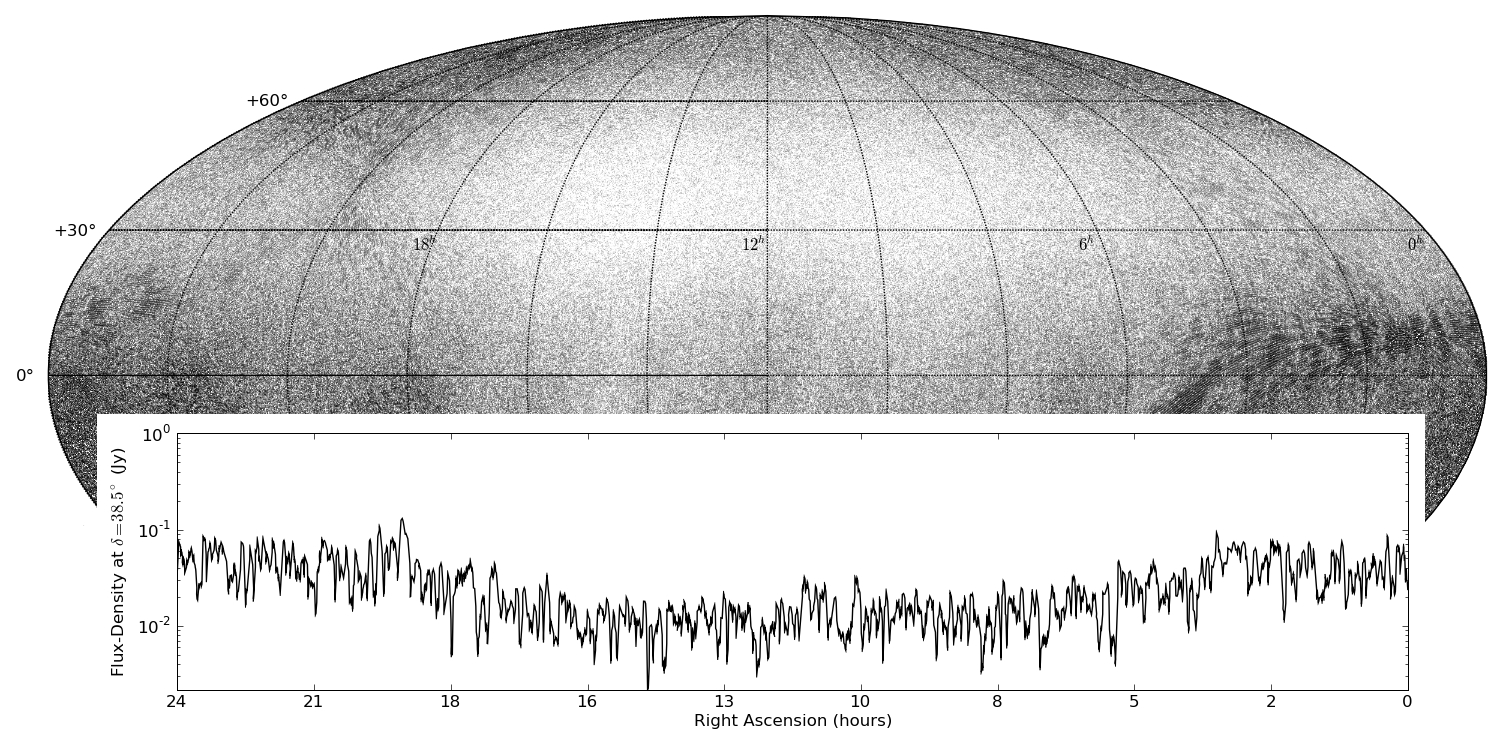}
    \caption{
Plotted above (top, gray-scale) is a noise map obtained by imaging the
PGB-8 data used to produce Figure \ref{fig:allsky_map}, but with consecutive
integrations added with alternating signs.  The map is in units of ${\rm
log}_{10}(Jy/Beam)$ ranging from $-2$ (white) to $0$ (black).
Flux densities in this map are indicative of thermal noise in the
northern hemisphere of Figure \ref{fig:allsky_map}.
The lower line-graph illustrates a the cut through the map at
declination $+38.5^\circ$, where PGB-8 is optimally sensitive. Flux densities in
this cut range from 10 mJy to 50 mJy, corresponding to temperatures of 620 mK
and 3.1 K at 156.4 MHz, using a synthesized beam size of 2.15e-5 steradians.
    }
    \label{fig:nos_map}
\end{sidewaysfigure*}

The angular power spectra illustrated in Figure \ref{fig:Cl_vs_l} are measured
using the PGB-8 data described in \S\ref{sec:results}.  Our analysis
follows the techniques developed in \citet{white_et_al1999} for calculating
angular power spectra with interferometric data in the flat-field
approximation.  To make this approximation, gridded visibility data projected
towards each of the four phase centers defined in \S\ref{sec:px_noise} were
limited to a $10^\circ$ diameter FoV by applying a circular boxcar windowing
function $W$, whose effect in the $uv$-plane is to convolve by the
complementary Fourier kernel $\tilde{W}(\vec u)$.  Within each facet, angular
power spectra were calculated using the equation:
\begin{align}
    C_\ell&=\frac{1}{2\ell + 1}\sum_m{\left|a_{\ell m}\right|^2}\nonumber\\
    C(u)&\simeq
    \left(\frac{\lambda^2}{2k_{\rm B}\Omega_{\rm W}}\right)^2
    \frac{
    \oint{\left|V(\vec u)*\tilde{W}(\vec u)\right|^2 d^2u}}
    {2\pi u},
    \label{eq:flat_sky}
\end{align}

where $\Omega_{\rm W}$ represents the solid angle spanned by $W$; and $V(\vec
u)$ represents the measured visibility as a function of position $\vec
u\equiv(u,v)$ in the $uv$-plane; $\oint$ denotes integration around a 
ring of $|\vec
u|=u$.  The relationship between $uv$-coordinates and angular wavenumber is
defined by $\ell\simeq2\pi\left|\vec u\right|$.  In practice, $V(\vec u)$ is
accompanied by sample weights $B(\vec u)$, so that Equation \ref{eq:flat_sky}
becomes:
\begin{equation}
    C(u)\simeq
    \left(\frac{\lambda^2}{2k_{\rm B}\Omega_{\rm B}}\right)^2
    \frac{\oint{\left|\left(B(\vec u)V(\vec u)\right)
        *\tilde{W}(\vec u)\right|^2 d^2u}}
         {\oint{\left|B(\vec u)*\tilde{W}(\vec u)\right|^2 d^2u}},
\label{eq:actual_pspec}
\end{equation}
where $\Omega_{\rm B}$ represents the solid angle spanned by the primary beam.

For each facet, separate $C(u)$ spectra were formed in 1.46-MHz intervals by
integrating around rings in the $uv$-plane that were logarithmically spaced in 
radius and
width.  Within each ring, data were weighted by the square of the primary beam
response toward phase center for optimal SNR.  We estimated the contribution 
of thermal
noise to the error in each $C(u)$ measurement by generating power
spectra for thermal noise images.  This was done using the technique of
adding integrations with alternating signs described in the previous section.
The error from cosmic variance was much larger than the error from thermal
noise, owing to the limited aperture sampling of the PGB-8 array and the
limited area on the sky employed for this analysis.  Error contributions from
cosmic variance were estimated by measuring the standard deviation of all
$C(u)$ measurements in rings of constant $u$ in the $uv$-plane for each facet,
and dividing this by the square root of the number of independent $C(u)$
measurements made.  We estimated independent measurements 
by calculating the size of a coherence patch in the $uv$-plane that arises from
$\tilde{W}(\vec u)$ and then degrading the pixel resolution in the $uv$-plane
to a corresponding size. At the lower resolution, each pixel sample was
considered an independent measurement, with the caveat that for a real-valued
sky, only half of the samples in a $|\vec u|=u$ ring are independent.

PGB-8 angular power spectra are dominated by unclustered point
sources, as indicated by the approximate $\ell^2$ power-law in
$\ell(\ell+1)C_\ell/2\pi$ of the angular power spectra plotted in Figure
\ref{fig:Cl_vs_l}.  However, at lower radio frequencies for $\ell<100$, we see 
a flattening of these spectra to a shallow, negative power-law that shows
qualitative agreement with the $C_{\ell,{\rm sync}}\propto\ell^{-2.4}$ model
adopted by \citet{tegmark_et_al2000} and \citet{santos_et_al2005}.
Progress in modeling point sources
and in subtracting spectrally smooth emission associated with synchrotron
sources will facilitate the suppression of these foregrounds, with a
fundamental limit placed by the thermal limit of these observations (Fig.
\ref{fig:Cl_vs_l}, dot-dashed), which we measure to be 310 mK for $\ell=100$ at
160 MHz.  Overall, our measurements indicate that over the range
$10^2<\ell<10^3$, the fundamental sensitivity level of these observations is
between two and three orders of magnitude in temperature above the fiducial model
of reionization at $z=9.2$ ($\nu=$140 MHz) used by \citet{santos_et_al2005}.

\section{Conclusion}

PAPER is following an incremental build-out strategy that emphasizes
the characterization and optimization of instrumental performance at each
stage of development.
We have given special attention to designing antenna elements whose smooth
spatial and spectral responses are conducive to calibration that will be 
required to 
model and remove strong foregrounds to the 21cm EoR signal.
Facilitating the incremental build-out of PAPER is a series of
digital correlators that take advantage of the scalability and reusability of
CASPER's packet-switching correlator architecture.  We are developing
post-correlation
calibration, imaging, and analysis pipelines as part of the
open-source AIPY software package.  This package emphasizes the needs of 
low-frequency
interferometry and facilitates experimentation with new analysis architectures
and calibration techniques.

With results from the PWA-4 and PGB-8 deployments in Western
Australia and Green Bank, respectively, we demonstrate a first level of
calibration that is sufficient for modeling strong sources to a level where
ionospheric refraction and temperature-dependent gains must be taken into
account.  This calibration relies heavily on a static model of the primary
beam.  Next steps in calibration will focus on the first-order removal of
temperature-dependence in antenna gains and on measuring deviations of the
primary beam of each antenna from the computed model.

Data from the PWA-4 and PGB-8 deployments have been used to generate an all-sky
map that attains a thermal noise level of 10 mJy/Beam (corresponding to 620 mK,
for a 2.15e-5 steradian synthesized beam size at 154.4 MHz) integrated across a
138.8-MHz to 174.0-MHz band.  This achievement represents a first iteration in
a cycle of improvement wherein sky models are used to improve array
calibration, which in turn allows increasingly accurate
sky models to be generated.  In analyzing the noise characteristics of this map, we find the
system temperature of the PGB-8 array to be consistent with a model of galactic
synchrotron emission.  Angular power spectra
generated from these data indicate that point sources are currently the
dominant foreground to the EoR signature, with evidence for contributions from
galactic synchrotron emission at lower radio frequencies and angular wavemodes. 
Errors in $C_\ell$
measurements are dominated by cosmic variance, but by differencing
integrations, we measure the thermal noise level at $\ell=100$ and 160 MHz to
be 310 mK--a sensitivity threshold that lies between two and three orders of
magnitude in temperature above fiducial reionization models.
These results demonstrate the need for a next level of
calibration, modeling, and foreground suppression that will be pursued in the
next PAPER deployments.

This work was supported by NSF AST grants 0804508, 0505354, and 0607838, and by
significant efforts by staff at NRAO's Green Bank and Charlottesville sites.
NEG acknowledges support from the Virginia Space Grant Consortium.  JRM
acknowledges financial support from the MeerKAT project and South Africa's
National Research Foundation.  We thank R. Beresford \& T. Sweetnam of CSIRO
for their aid in the PWA-4 deployment and J. Richards and the Western
Australia government for their support.  CASPER research, including components
of PAPER's correlator development, is supported by the NSF grant AST-0619596.
We would like to acknowledge the students, faculty and sponsors of the Berkeley
Wireless Research Center, and the NSF Infrastructure grant 0403427. 


\bibliographystyle{apj}
\bibliography{biblio}

\begin{thebibliography}{38}
\expandafter\ifx\csname natexlab\endcsname\relax\def\natexlab#1{#1}\fi

\bibitem[{{Baars} {et~al.}(1977){Baars}, {Genzel}, {Pauliny-Toth}, \&
  {Witzel}}]{baars_et_al1977}
{Baars}, J.~W.~M., et al.  1977, \aap, 61, 99

\bibitem[{{Barkana} \& {Loeb}(2005{\natexlab{a}})}]{barkana_loeb2005a}
{Barkana}, R. \& {Loeb}, A. 2005{\natexlab{a}}, \apjl, 624, L65

\bibitem[{{Barkana} \& {Loeb}(2005{\natexlab{b}})}]{barkana_loeb2005b}
---. 2005{\natexlab{b}}, \apj, 626, 1

\bibitem[{{Becker} {et~al.}(2001){Becker}, {Fan}, {White}, {Strauss},
  {Narayanan}, {Lupton}, {Gunn}, {Annis}, {Bahcall}, {Brinkmann}, {Connolly},
  {Csabai}, {Czarapata}, {Doi}, {Heckman}, {Hennessy}, {Ivezi{\'c}}, {Knapp},
  {Lamb}, {McKay}, {Munn}, {Nash}, {Nichol}, {Pier}, {Richards}, {Schneider},
  {Stoughton}, {Szalay}, {Thakar}, \& {York}}]{becker_et_al2001}
{Becker}, R.~H., et al. 2001, \aj, 122, 2850

\bibitem[{{Bowman} {et~al.}(2007){Bowman}, {Barnes}, {Briggs}, {Corey},
  {Lynch}, {Bhat}, {Cappallo}, {Doeleman}, {Fanous}, {Herne}, {Hewitt},
  {Johnston}, {Kasper}, {Kocz}, {Kratzenberg}, {Lonsdale}, {Morales}, {Oberoi},
  {Salah}, {Stansby}, {Stevens}, {Torr}, {Wayth}, {Webster}, \&
  {Wyithe}}]{bowman_et_al2007a}
{Bowman}, J.~D., et al. 2007, \aj, 133, 1505

\bibitem[{{Bowman} {et~al.}(2006){Bowman}, {Morales}, \&
  {Hewitt}}]{bowman_et_al2006}
{Bowman}, J.~D., et al. 2006, \apj, 638, 20

\bibitem[{{Bradley}(2006)}]{bradley2006}
{Bradley}, R. 2006, {A Low Cost Screened Enclosure for Effective Control of
  Undesired Radio Frequency Emissions}, NRAO EDIR 317,
  http://www.gb.nrao.edu/electronics/edir

\bibitem[{{Cornwell} \& {Fomalont}(1989)}]{cornwell_fomalont1989}
{Cornwell}, T. \& {Fomalont}, E.~B. 1989, in ASPCS,
  Vol.~6, Synthesis Imaging in Radio Astronomy, ed.
  R.~A. {Perley}, F.~R. {Schwab}, \& A.~H. {Bridle}, 185--+

\bibitem[{{Cornwell} \& {Evans}(1985)}]{cornwell_evans1985}
{Cornwell}, T.~J. \& {Evans}, K.~F. 1985, \aap, 143, 77

\bibitem[{{Cornwell} {et~al.}(2003){Cornwell}, {Golap}, \&
  {Bhatnagar}}]{cornwell_et_al2003}
{Cornwell}, T.~J., et al. 2003, {W-Projection: A New
  Algorithm for Non-Coplanar Baselines}, EVLA Memo~67

\bibitem[{{Crochiere} \& {Rabiner}(1983)}]{crochiere_rabiner1983}
{Crochiere}, R. \& {Rabiner}, L.~R. 1983, {Multirate Digital Signal Processing}
  (Englewood Cliffs, NJ, Prentice-Hall, 1983.~336 p.)

\bibitem[{{Fan} {et~al.}(2006){Fan}, {Carilli}, \& {Keating}}]{fan_et_al2006}
{Fan}, X., {Carilli}, C.~L., \& {Keating}, B. 2006, \araa, 44, 415

\bibitem[{{Furlanetto} {et~al.}(2006){Furlanetto}, {Oh}, \&
  {Briggs}}]{furlanetto_et_al2006}
{Furlanetto}, S.~R., et al. 2006, \physrep, 433, 181

\bibitem[{{G{\'o}rski} {et~al.}(2005){G{\'o}rski}, {Hivon}, {Banday},
  {Wandelt}, {Hansen}, {Reinecke}, \& {Bartelmann}}]{gorski_et_al2005}
{G{\'o}rski}, et al. 2005, \apj, 622, 759

\bibitem[{{Hall}(1988)}]{arrl1988}
{Hall}, G., ed. 1988, {The ARRL Antenna Book, 15th Edition} (American Radio
  Relay League, Newington, CT, 1988.), 16

\bibitem[{{Haslam} {et~al.}(1982){Haslam}, {Salter}, {Stoffel}, \&
  {Wilson}}]{haslam_et_al1982}
{Haslam}, C.~G.~T., et al. 1982, \aaps, 47, 1

\bibitem[{{H{\"o}gbom}(1974)}]{hogbom1974}
{H{\"o}gbom}, J.~A. 1974, \aaps, 15, 417

\bibitem[{{Jackson} \& {Farrell}(2006)}]{jackson_farrell2006}
{Jackson}, T.~L. \& {Farrell}, W.~M. 2006, IPPS, 44, 2942

\bibitem[{{Johnson}(1993)}]{johnson1993}
{Johnson}, R.~C. 1993, {Antenna Engineering Handbook, 3rd Ed.} (New York,
  McGraw-Hill, 1993.), 18--23

\bibitem[{{Komesaroff}(1960)}]{komesaroff1960}
{Komesaroff}, M.~M. 1960, AuJP, 13, 153

\bibitem[{{Loeb} \& {Barkana}(2001)}]{loeb_barkana2001}
{Loeb}, A. \& {Barkana}, R. 2001, \araa, 39, 19

\bibitem[{{Loeb} \& {Zaldarriaga}(2004)}]{loeb_zaldarriaga2004}
{Loeb}, A. \& {Zaldarriaga}, M. 2004, PRL, 92, 211301

\bibitem[{{Morales}(2005)}]{morales2005}
{Morales}, M.~F. 2005, \apj, 619, 678

\bibitem[{{Morales} {et~al.}(2006){Morales}, {Bowman}, {Cappallo}, {Hewitt}, \&
  {Lonsdale}}]{morales_et_al2006b}
{Morales}, M.~F., et al. 2006, NewAR, 50, 173

\bibitem[{{Page} {et~al.}(2007){Page}, {Hinshaw}, {Komatsu}, {Nolta},
  {Spergel}, {Bennett}, {Barnes}, {Bean}, {Dor{\'e}}, {Dunkley}, {Halpern},
  {Hill}, {Jarosik}, {Kogut}, {Limon}, {Meyer}, {Odegard}, {Peiris}, {Tucker},
  {Verde}, {Weiland}, {Wollack}, \& {Wright}}]{page_et_al2007}
{Page}, L., et al. 2007, \apjs, 170, 335

\bibitem[{{Parsons} {et~al.}(2008){Parsons}, {Backer}, {Siemion}, {Chen},
  {Werthimer}, {Droz}, {Filiba}, {Manley}, {McMahon}, {Parsa}, {MacMahon}, \&
  {Wright}}]{parsons_et_al2008}
{Parsons}, A., et al. 2008, \pasp, 120, 1207

\bibitem[{{Parsons} \& {Backer}(2009)}]{parsons_backer2009}
{Parsons}, A.~R. \& {Backer}, D.~C. 2009, Sub. to \aj. ArXiv:0901.2575

\bibitem[{{Rogers} \& {Bowman}(2008)}]{rogers_bowman2008}
{Rogers}, A.~E.~E. \& {Bowman}, J.~D. 2008, \aj, 136, 641

\bibitem[{{Santos} {et~al.}(2005){Santos}, {Cooray}, \&
  {Knox}}]{santos_et_al2005}
{Santos}, M.~G., et al. 2005, \apj, 625, 575

\bibitem[{{Sault}(1990)}]{sault1990}
{Sault}, R.~J. 1990, \apjl, 354, L61

\bibitem[{{Sault} {et~al.}(1995){Sault}, {Teuben}, \&
  {Wright}}]{sault_et_al1995}
{Sault}, R.~J., et al. 1995, in ASPCS,
  Vol.~77, ADASS IV, ed. R.~A. {Shaw}, H.~E. {Payne}, \& J.~J.~E.
  {Hayes}, 433--+

\bibitem[{{Slurzberg} \& {Osterheld}(1961)}]{slurzberg_osterheld1961}
{Slurzberg}, M. \& {Osterheld}, W. 1961, {Essentials of Radio-Electronics, 2nd
  Ed.} (New York, McGraw-Hill, 1961.), 595

\bibitem[{{Tegmark} {et~al.}(2000){Tegmark}, {Eisenstein}, {Hu}, \& {de
  Oliveira-Costa}}]{tegmark_et_al2000}
{Tegmark}, M., et al. 2000, \apj, 530, 133

\bibitem[{{Thompson} {et~al.}(2001){Thompson}, {Moran}, \&
  {Swenson}}]{thompson_et_al2001}
{Thompson}, A.~R., et al. 2001,
  {Interferometry and Synthesis in Radio Astronomy, 2nd Edition} (New York,
  Wiley-Interscience, 2001.~692 p.)

\bibitem[{{Vaidyanathan}(1990)}]{vaidyanathan1990}
{Vaidyanathan}, P.~P. 1990, Proc. IEEE, 78, 56

\bibitem[{{White} {et~al.}(1999){White}, {Carlstrom}, {Dragovan}, \&
  {Holzapfel}}]{white_et_al1999}
{White}, M., et al. 1999, \apj, 514, 12

\bibitem[{{Yatawatta} {et~al.}(2008){Yatawatta}, {Zaroubi}, {de Bruyn},
  {Koopmans}, \& {Noordam}}]{yatawatta_et_al2008}
{Yatawatta}, S., et al. 2008, ArXiv:0810.5751

\bibitem[{{Zahn} {et~al.}(2007){Zahn}, {Lidz}, {McQuinn}, {Dutta}, {Hernquist},
  {Zaldarriaga}, \& {Furlanetto}}]{zahn_et_al2007}
{Zahn}, O., et al. 2007, \apj, 654, 12

\end{thebibliography}


\end{document}